\documentclass[aip,preprint,amssymb, amsmath,onecolumn]{revtex4-1}
\usepackage[utf8]{inputenc}
\usepackage{bm}%

\usepackage{natbib}
\usepackage{graphicx}
\usepackage{color}
\usepackage{verse}

\begin{document}


\title{The fluid mechanics of bubbly drinks}

\author{Roberto Zenit}
\email{zenit@unam.mx}
\affiliation{Instituto de Investigaciones en Materiales, Universidad Nacional Aut\'onoma de M\'exico, Ciudad de M\'exico, 04510, M\'exico}

\author{Javier Rodr\'{\i}guez-Rodr\'{\i}guez}%
\email{bubbles@ing.uc3m.es}
\affiliation{Fluid Mechanics Group, Universidad Carlos III de Madrid, Legan\'es, 28911, Spain}

\date{\today}

\begin{abstract}
Bubbly drinks are surprisingly attractive. There is something about the nature of the these beverages that make them preferable among other choices. In this article we explore the physics involved in this particular kind of two-phase, mass-transfer-driven flows.
\end{abstract}

\maketitle

\begin{verse}[\versewidth]
Tiny bubbles \\
In the wine \\
Make me feel happy \\
Ah, they make me feel fine ...\\
\begin{flushright}
Don Ho, American musician, 1966. \\
\end{flushright}
\end{verse}

\section{Why bubbly?}

Most people find bubbly drinks to be attractive and refreshing. In most cases, these bubbles result from the carbonation of the water that serves as main ingredient to the drink. When water is exposed to carbon dioxide, CO$_2$, the latter dissolves into the liquid in an amount that is proportional to the pressure, with a proportionality constant such that the amount of CO$_2$ dissolved  is large at typical bottling pressures. If the pressure is then suddenly reduced, as happens when a bottle is opened, the gas comes out of solution rapidly, forming many bubbles that rise to the surface. These surface bubbles either burst after a brief instant or are able to remain floating and aggregate to form a foam head. So, as you see, many physical processes occur in your glass before your first refreshing sip. In this article we address some of them. Motivated by our love for bubbly drinks, we use them as a vehicle to discuss the physics of the bubble formation, motion and stability, and to explore the fascinating connections with a range of other phenomena that lay far beyond the need to cool one self.

Perhaps the first question that needs to be asked is: why do we like bubbly drinks? Several studies have considered this very issue in the past. It has been found that carbonation, in fact, triggers pain receptors in the deep brain such as those activated when tasting spicy food \cite{Wang2010}. Interestingly, when given to other animals such as mice, dogs or horses, they refuse to drink carbonated water. Humans for a strange reason like these mildly irritating effects. Bubbles, of course, play a central role in the perception of carbonation. To begin with, when CO$_2$ bubbles appear in water, and in combination with some enzymes found in saliva, a reaction occurs that leads to the formation of small amounts of carbonic acid, the substance believed to be behind the tingly sensation of bubbly drinks.  It is known also that the size of the bubbles alters the sensory perception of a given carbonated drink. For example, having small bubbles would favor the transfer of mass and, in consequence, increase the production of carbonic acid. For the case of soda (water with sugar), the amount and size of bubbles also affects the perception of flavor. In fact, manufacturers usually adjust the gas content for different drinks to please the consumers palate (sweet drinks usually have more gas than plain ones). To investigate the effect bubbles have in the perception of flavor, some studies \cite{Wise2013} have tested the role of the bubble size distribution in the taste of carbonated liquids. Puzzling results were found: on the one hand, bubbles were not required to experience the carbonation bite but, on the other hand, they did modulate the perceived flavor. Clearly, we still do not have clear picture of the actual mechanism by which bubbles influence taste.

Carbonation can occur naturally or artificially. The artificial process of carbonation was invented by, non-other than, Joseph Priestley, better known for the discovery of Oxygen. In 1772 he discovered that air could be dissolved into water at high pressures. The original intention was to serve as a method to maintain water potable in ships. Even then, the most relevant result was the ``distinct freshness'' of drinking the bubbly liquid.

Naturally occurring carbonation is, essentially, the result of the same process. The sparkling water from the French town of Verg\`eze (where the commercial brand Perrier is bottled) is naturally carbonated since the underwater water source is at high pressure and exposed to a natural source of carbon dioxide. The other more interesting way to naturally carbonate a drink is that resulting from fermentation. When yeast eats simple sugars it excretes,  mostly, ethanol and carbon dioxide. If the process occurs in a closed container,  the pressure rises as the amount of carbon dioxide increases. In turn, as the pressure increases, the gas can dissolve more readily into the liquid. Although beer making can be dated back for thousands of years, being one of the oldest prepared beverages \cite{Dietrich2012}, it is unclear how bubbly this ancient beer could have been. The ceramic containers would have to be sealed and be able to sustain pressures up to a few atmospheres. Sparkling wine was discovered later, in the 17th century. In it the carbonation results from a secondary fermentation that occurs within the bottle itself. The creation of Champagne has been credited to Dom Perignon, the wine-master in a French abbey but some studies indicate that sparkling wine could have existed as early the XIV century. Perignon, however, did invent strong bottles and special corks to  prevent the bottles from exploding during the process. This delicacy would not become popular for more than 100 years after it was first invented. 

The presence of alcohol and other by-products of fermentation, such as proteins and enzymes, in bubbly drinks makes the physical description even more interesting. The physical properties of the liquid are significantly affected which, in turn, affects the  bubble formation and motion, as well as its surface stability. No less important is the effect of alcohol ingestion when drinking a bubbly liquid of this type. As discussed above, gas bubbles affect the sensation and perception of flavor but they also accelerate the absorption of alcohol in the body\cite{Ridout2003}. The human consumption of alcoholic drinks is as long as civilization itself. Alcohol consumption has many social and cultural implications. 
Despite variations of different cultures, in most cases drinking alcoholic beverages is associated with celebration; it is a sort of universal symbol of festivity, social bonding and integration. No other substance possesses this status in human life.

\begin{figure}[h!]
\centering
\includegraphics[width=0.9\textwidth]{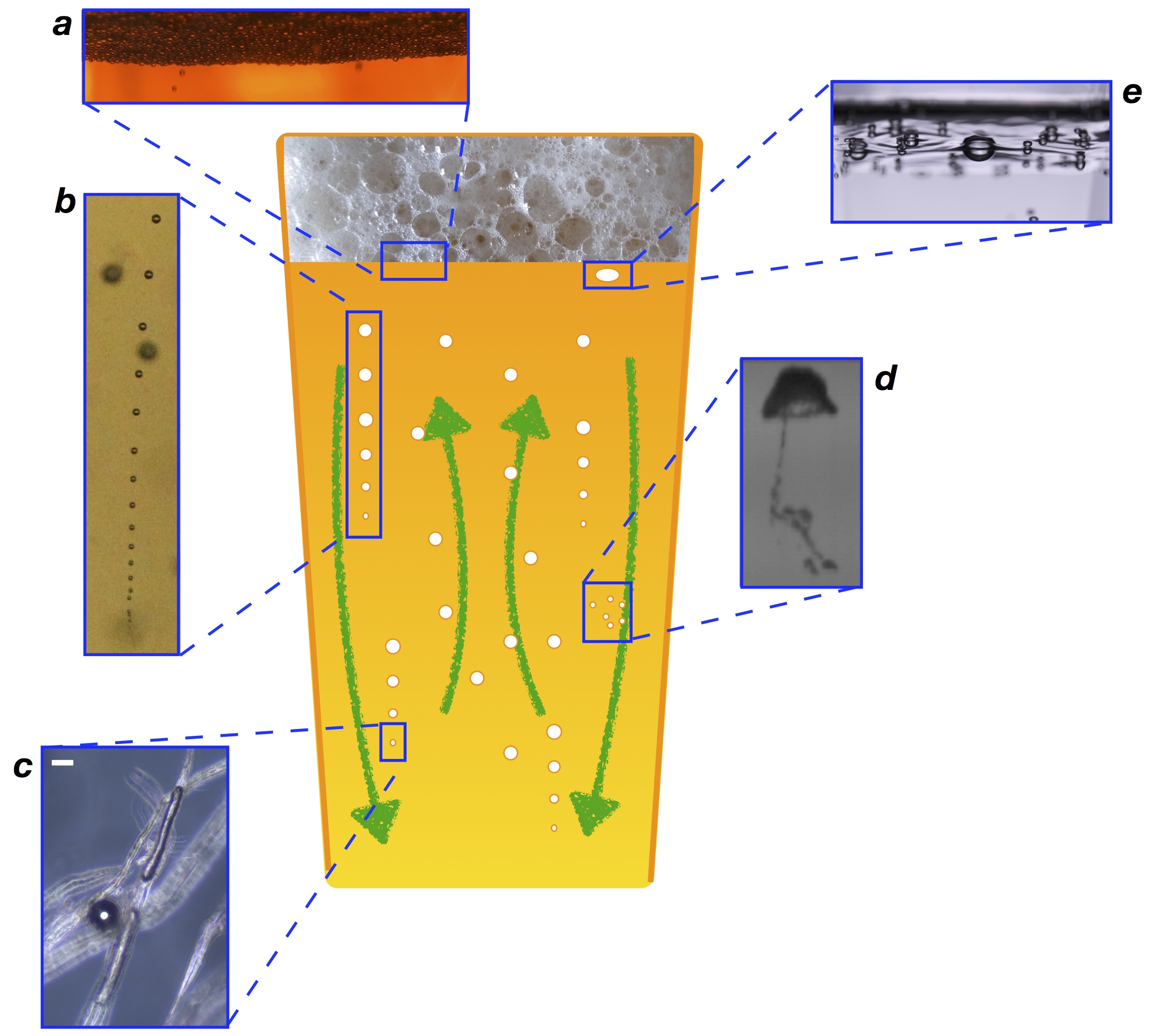}
\caption{\footnotesize{Sketch of a glass of a bubbly drink showing the different phenomena that we treat in this article. ($a$) Head of foam formed on the surface of a beer glass. ($b$) Trail of bubbles that rise from a nucleation site. It can be clearly seen how bubbles grow as they rise, due to mass transfer. The rising of bubbles from different nucleation sites along the glass induces a global circulation (sketched here with green arrows) that greatly enhances the degassing of the drink. This circulation could, in fact, be in either the direction shown in the figure or the opposite one, depending on whether bubbles appear mostly at the walls, around the center of the glass or homogeneously everywhere. ($c$) Cellulose fibers in beer that serve as bubble nucleation sites. The gas cavities inside the fiber are clearly visible. One bubble has just come out of the fiber. The scale bar is 50 $\mu$m. ($d$) Mushroom-shaped bubble plume arising from the implosion of a millimetric bubble when a beer bottle is tapped on its mouth. The head of the mushroom is around 3 mm wide. ($e$) If surface bubbles are not stable enough, they burst before they can aggregate to form foam. See Appendix for photo credits.}}
\label{fig:bubbly_drink}
\end{figure}

Alcoholic or not, bubbly drinks are full of physics. In figure \ref{fig:bubbly_drink} we show a schematic view of the most relevant processes that take place when a carbonated drink is poured into a tall glass (see Movie 1, in the Supplementary Material). If the liquid is poured into the glass shortly after the pressurized bottle was opened, the `birth' of bubbles will be observed both in the bulk liquid and over the surface of the glass (see insert ($c$) in figure \ref{fig:bubbly_drink} and section \ref{sec:birth_rise} below). The continuous formation of streams of bubbles will induce a net motion of the liquid inside the glass, a sort of bubble-driven convection (see insert ($b$) in figure \ref{fig:bubbly_drink}), that will affect the bubble production rate and motion. As they grow, bubbles will rise and eventually reach the surface. Once reaching the surface, and depending on the properties of the liquid, the bubbles will either burst or float on the surface for some time (figure \ref{fig:bubbly_drink}$e$). If the life time of floating bubbles is large then the continued arrival of other bubbles will lead to the formation of a foam head (figure \ref{fig:bubbly_drink}$a$). In what follows, we discuss each of these processes in more detail.

\newpage

\section{\label{sec:birth_rise}The birth and early life of bubbles: nucleation, growth and rise}

The behavior of bubbles in carbonated drinks is mainly determined by two physico-chemical laws. The first one is Henry's law: the concentration of dissolved gas at a liquid-gas interface, $C_s$, is proportional to the partial pressure of the gas, $p_g$. Mathematically,
\begin{equation}
C_s = k_H p_g,
\end{equation}
where $k_H$ is the so-called Henry's constant (although it actually depends on the temperature). $C_s$ is usually termed the saturation concentration, as it represents how much gas can be dissolved in the liquid at a given pressure. This equation has an important implication: if a liquid has been pressurized with a soluble gas, say CO$_2$, for sufficient time, it will contain more gas in solution that what is able to store when exposed to the (lower) atmospheric pressure. We say then that the liquid is supersaturated at ambient pressure. Typical bottling pressure for beer is about 3 atm, whereas champagne is usually bottled at 6 atm. Since Henry's constant for CO$_2$ in water at 12$^{\rm o}$C is $k_H = 1.9$ g/L$\cdot$atm, this means that a beer bottle contains between 5-6 g/L whereas a champagne one contains about 11 g/L. If these CO$_2$ gas masses were kept at normal conditions, they would occupy 3 and 5.6 liters respectively. Compare these figures with the gas content that corresponds to saturation at ambient pressure, $C_s \approx 1.9$ g/L, which would occupy only 1 liter.

A second one is Fick's law of molecular diffusion, analogous to Fourier's law of heat conduction. This law states that in the presence of a concentration gradient a mass flux proportional to this gradient, but opposed to it, is established. The proportionality constant is the diffusivity of CO$_2$ in water, $D \approx 2\times 10^{-9}$ m$^2$/s.

With these tools now at hand we can take a look at how bubbles are born in the first place. When a liquid that has been bottled with pressurized CO$_2$ is exposed to the ambient pressure, Henry's law dictates that the interface of any tiny gas cavity present in the bottle will immediately adopt the saturation concentration corresponding to the ambient pressure, which is lower than that of the bulk liquid. In response to this concentration gradient, molecular diffusion will induce a net mass flux of gas towards the cavity which will make it grow.

%
%
\begin{center}
\colorbox{cyan}{\parbox{0.9\textwidth}{{\bf BOX 1. The Epstein--Plesset equation}\\
For the case of a spherical isolated bubble Henry's law and Fick's diffusion combine to yield the so-called Epstein--Plesset equation \citep{EpsteinPlesset1950}, which serves as a building block for more elaborated models of bubble-liquid mass exchange. If surface tension is not considered (it is only important for bubbles with a radius smaller than about $R_c \approx 2\sigma/P_0$, with $\sigma$ the surface tension coefficient, with yields $R_c \approx 1$ $\mu$m), this equation reads:
\begin{equation*}
    \frac{\mathrm{d}R}{\mathrm{d}t} = D\Lambda\zeta\left(\frac{1}{R} + \frac{1}{\sqrt{\pi D t}}\right).
\end{equation*}
Here, $\Lambda = k_H R^0 T$ is the Ostwald constant, with $R^0$ the universal gas constant and $T$ the temperature; and $\zeta = P_b/P_0 - 1$ is the supersaturation, with $P_b$ and $P_0$ the bottling and ambient pressure respectively.
This equation predicts that the size of a gas bubble grows by pure diffusion as the square root of the time. Professor Detlef Lohse, from the University of Twente, described this as the ``Hydrogen atom'' of mass transfer processes involving bubbles (and drops), since more complex theories build on it very much like molecular physics builds on the solution of the Schr\"odinger equation for the Hydrogen atom.}}\\
\end{center}
%
%

These gas cavities are formed before the filling and survive either by attaching on crevices at the walls, which stabilizes them by pinning, or thanks to impurities able to keep some gas content inside, such as cellulose fibers \cite{LigerBelair2005} (see figure \ref{fig:bubbly_drink}(c) and also Movie 2 in the supplementary material). Although in principle the nucleation could start at any imperfection of the walls even without the previous existence of a gas nucleation site, the level of supersaturation needed for this to be feasible is too large and would rarely be observed. 

When the gas cavities reach a size large enough for buoyancy to detach them from the nucleation site, they start to rise, albeit usually leaving behind some smaller gas cavity that will make the cycle start over again. This is the origin of the beautiful bubble trails commonly observed in beer or champagne glasses (see panel ($b$) in figure \ref{fig:bubbly_drink}). New bubbles join the trail from the nucleation site at a frequency given by the cycle described above. Once in the trail, they rise with a speed that grows with their radius, which explains their increasing separation distance.

\begin{figure}
    \centering
    \begin{tabular}{cc}
    \includegraphics[height=0.37\textwidth]{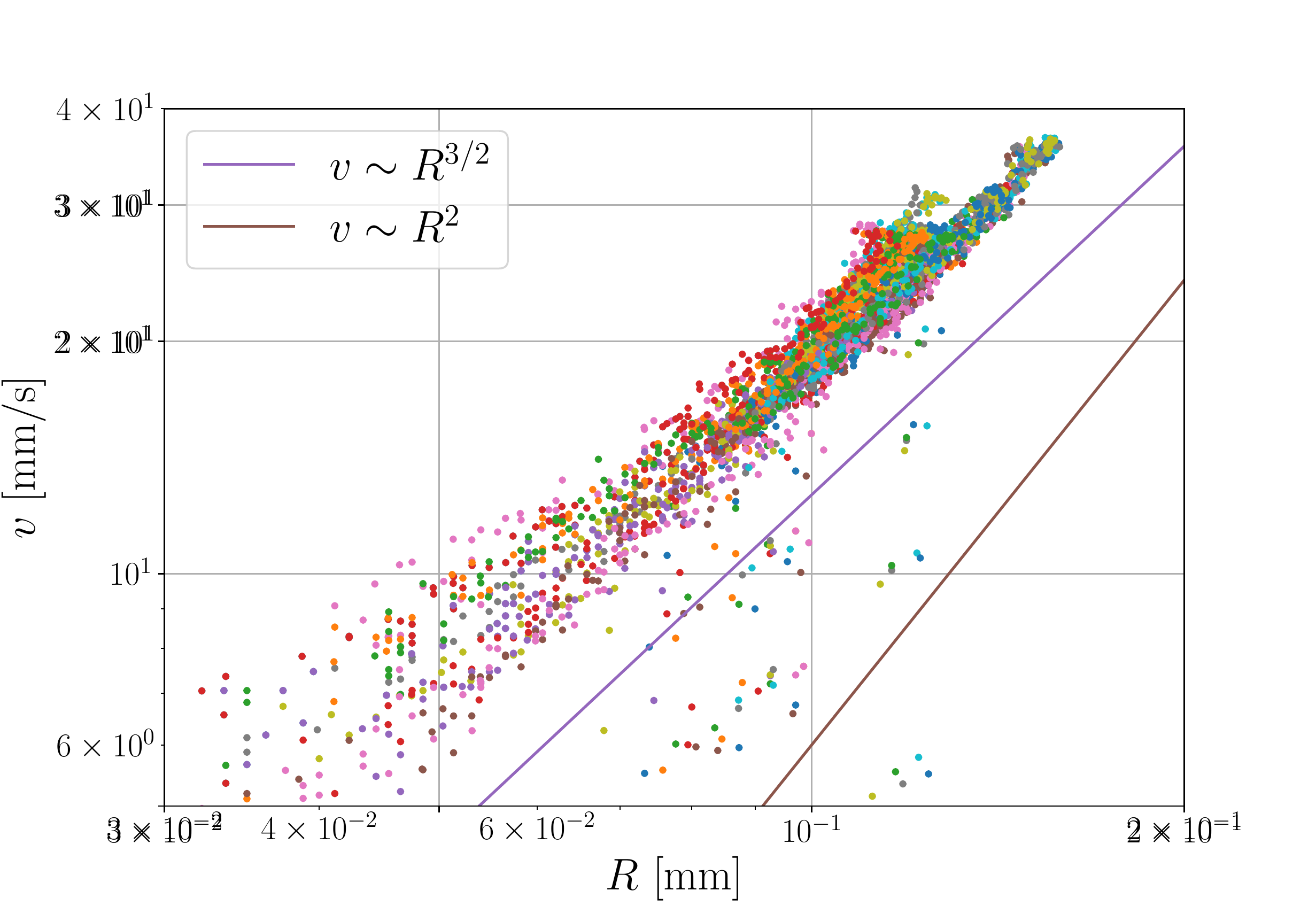} &
    \includegraphics[height=0.37\textwidth]{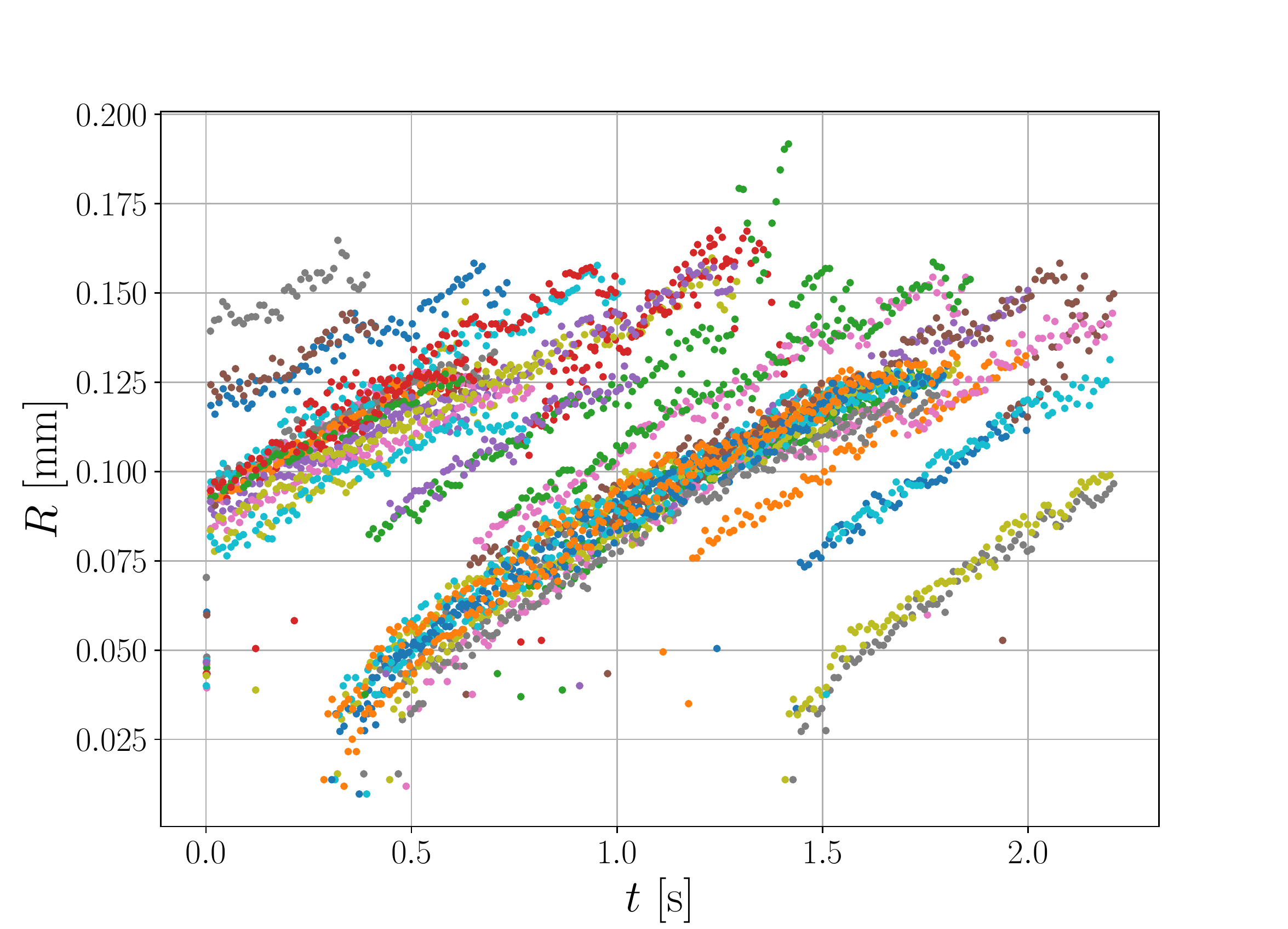}\\
    ($a$) & ($b$)
    \end{tabular}
    \caption{($a$) Rising velocity of CO$_2$ bubbles in beer. Bubbles are produced in the bulk liquid by a focused laser pulse to prevent any effect of nearby walls. There exists a power law relation between the bubble rising velocity and its radius, although the exponent seems to be closer to 3/2 than to 2, due to the finite values of the Reynolds numbers, specially for the largest bubbles (the Reynolds number ranges from 0.35 to about 12). ($b$) Time evolution the radii of the same bubbles shown in panel $a$. Except for very small sizes, or conversely times, where the growth is mostly driven by diffusion, the bubble growth rate is fairly constant for all the bubble regardless of its size. Experiments were done in a lager beer with a bottling pressure of 5 bars.}
    \label{fig:bubble_rise}
\end{figure}

Naturally, when a bubble rises in a quiescent liquid the bubble-liquid relative velocity enhances mass transfer and the bubble grows faster than the scaling $R \sim t^{1/2}$ predicted by the Epstein--Plesset equation (see Box 1). Indeed, the radius of a bubble that travels upwards in a supersaturated liquid grows with time at a constant rate that is independent of its size\cite{LigerBelair2005}, as can be seen in figure \ref{fig:bubble_rise}b. The constancy of this growth rate can be understood in relatively simple terms using dimensionless numbers (for definitions of these numbers, see Box 2). The upwards motion of the bubble is driven by a balance between buoyancy, which scales with the bubble's volume, or cubed radius, and viscous drag. At the moderately small values of the Reynolds number at which bubbles commonly rise in beer or champagne (for instance, for bubbles shown in figure \ref{fig:bubble_rise}, the Reynolds ranges between about 0.35 and 12) the viscous drag is well approximated by Stokes' law, which states that, for a bubble of radius $R$ rising with velocity $v$ in a liquid of viscosity $\mu$, $F_D \sim \mu R v$. Thus, the balance between buoyancy and Stokes' drag suggests that the rising velocity is expected to scale as $v \sim g R^2/\nu$, with $\nu = \mu/\rho$ the kinematic viscosity, $\rho$ the liquid density and $g$ the gravitational acceleration. This means that both the Reynolds and Peclet numbers, which are proportional to the product $v R$, grow with $R^3$. On the other hand, because CO$_2$ diffuses in water 500 times slower than momentum, the Peclet number associated to this motion is large. In this regime of small $Re$ and large $Pe$, the Sherwood number, which in the dimensionless rate at which the bubble exchanges mass with the surrounding liquid, is nearly $Sh \approx Pe^{1/3}$. Combining the definitions of these parameters, we have
\begin{equation}
    \dot{R} \sim \Lambda\zeta\frac{D}{R}\left(\frac{g R^3}{D\nu}\right)^{1/3} \sim \Lambda\zeta(Dg)^{1/3}Sc^{-1/3},
\end{equation}
which is constant. Typically, this constant value of $\dot{R}$ is of the order of tenths of millimeters per second for bubbles found in drinks. Notice that this is the reason why bubbles grow as they rise (as shown in figure \ref{fig:bubbly_drink}$b$), not by the change in hydrostatic pressure, that would be only about 1\% of the ambient one for a container of 10 cm.
Another phenomenon that, in general, may alter the arguments above is the so-called virtual mass effect, i.e. the apparent inertia of a bubble with a time-varying radius. For bubbles in drinks this effect is small due to the moderate values of the rate of bubble growth, $\dot{R}$.

%
%
\begin{center}
\colorbox{cyan}{\parbox{0.9\textwidth}{{\bf Box 2. Relevant dimensionless numbers in the life of a bubble}\\
{\bf Reynolds number}, $Re = UR/\nu$, ratio between fluid inertia and viscous forces.\\
{\bf Peclet number}, $Pe = UR/D$, ratio between advective transport of a dissolved species and molecular diffusion.\\
{\bf Schmidt number}, $Sc = \nu/D$, ratio of viscous to molecular diffusivity. Note that $Pe = Sc Re$. Because for the case of CO$_2$ in water $Sc \approx 500$, the rise of a bubble in the glass occurs at a regime such that $Re$ is small but still $Pe$ is large.\\
{\bf Ostwald coefficient}, $\Lambda = k_H R^0 T$, volume of liquid that can be saturated with a given volume of gas. For CO$_2$ at 12$^o$C, $\Lambda \approx 1$, which means that the carbon dioxide dissolved in one liter of saturated water nearly occupies one liter if exolved from the solution.\\
{\bf Supersaturation level}, $\zeta = P_b/P_0 - 1$, dimensionless excess of gas dissolved compared to the one at saturation conditions.\\
{\bf Sherwood number}, $Sh = KR/D$, rate of mass transfer per unit surface and time divided with that due to pure molecular diffusion at long times. In the case of a bubble growing by mass diffusion, $K = \dot{R}/(\Lambda\zeta)$. In the regime where usually bubbles rise in a carbonated drink, and taking into account that their surfaces are populated with surfactants, $Sh \sim Pe^{1/3}$.\\
{\bf Bond number}, $Bo = \rho g R^2/\sigma$, compares the bubble size with the capillary length. It is of order $O(0.1)$ for a millimetric bubble in water.
}}\\
\end{center}
\vspace{0.5cm}

%
%

Similarly, a bubble sitting at a solid surface exhibits a faster growth if the liquid is flowing inside the container. Thus, there is a feedback between bubble growth and rise and the generation of a circulatory motion inside a glass. This circulatory motion induces a velocity near the walls that promotes faster growth. In turn, the rising motion of the bubbles provides the momentum needed to sustain the circulatory motion.  In other words, it is a self-sustained process that continues in time until a substantial part of the CO$_2$ dissolved initially is lost. This is the chain of events, in which bubbles are essential, that leads to the degassing of a beer, or champagne, glass left open to the ambient. Indeed, diffusion by itself would take an extremely long time to degas a glass of a carbonated drink. Let us illustrate this with an example. On dimensional grounds, it is possible to build a time scale for the time taken by diffusion to degas a container: $t_d \sim H^2/D$, with $H$ the height of the liquid. If we take $H = 10$ cm for a typical glass, this yields a timescale $t_d \approx 5\times10^6$ s, or about 1400 hours! Clearly diffusion alone cannot be held responsible for the lost of gas in the drink, some other effect may be into play.

Gerald Liger-Belair pointed out the role that bubbles have in solving this mystery\citep{LigerBelair2005}. Although the gas that they directly transport to the surface is only 20\% of the total volume, their rising motion creates a circulation inside the container that ultimately enhances the transport of CO$_2$ to the free surface by advection. In other words, they create the mixing necessary for the bulk liquid to transport the gas to the free surface as sketched in figure \ref{fig:bubbly_drink}. 

This set of phenomena that lead to the efficient (and unfortunate) degassing of a carbonated beverage can be dramatically sped up if a bottle is tapped on its mouth. A study in which one of us participated a few years ago \citep{RodriguezRodriguez_etalPRL2014} showed that an impact on the bottle triggers the formation of dense bubble clouds, consisting on up to million of microbubbles, due to cavitation. These bubble clouds are much more effective than individual bubbles in generating a convective motion inside the bottle. In fact, experiments revealed the existence of bubbly plumes, analogous to mushroom clouds forming upon an explosion, which have a substantial effect in promoting mixing and thus degassing (see figure \ref{fig:bubbly_drink}$d$). Similar phenomena are found in chemical reactors when the products of a reaction are lighter (or warmer) than the reactants. 

A thought-provoking idea is to connect these phenomena with others occurring in geology, namely gas-driven eruptions in lakes, also known as limnic eruptions \citep{ZhangKlingAREPS2006}. These eruptions are formed in lakes where, due to geological or biological activity, the bottom becomes supersaturated with CO$_2$. Because carbonated water is heavier than fresh water, this stratification is stable, which means that a lake can accumulate substantial amounts of carbon dioxide in its bottom. It occurs sometimes, for reasons not yet fully understood, that a fluid parcel from near the bottom rises to a shallower depth where it becomes supersaturated and therefore bubbles nucleate and grow. This then creates a plume that, if reaching the free surface, establishes a bubble-laden conduit that will continue degassing the bottom of the lake until the amount of dissolved CO$_2$ is too low to further support the plume. This is what happened in 1986 in Lake Nyos, Cameroon. As a result, 1700 people died of suffocation in the vicinity of the lake.

The importance of bubble growth and rise in a supersaturated liquid in geology goes beyond limnic eruptions. It is well accepted that the sudden exolution of volatile elements in magma during eruptions contributes to strengthen the intensity of certain kind of volcanic eruptions, through a mechanism similar to that occurring when Mentos are dropped in a bottle of Diet Coke, as pointed out by vulcanologists Katherine Cashman and Stephen Sparks \cite{CashmanSparks2013}. In view of these evidences, it is tempting to think that the autocatalytic nature of soluble bubble plumes like those appearing in beer tapping could be relevant to explain some geological phenomena not yet well understood. Most interestingly, autocatalytic bubbly plumes may play a role in the triggering of volcanic or geyser eruptions by distant seismic events. Although there are numerous evidences, gathered for more than a century, that suggest that volcanoes and geysers erupt in distant regions of the Earth in the days following a strong earthquake, the mechanism by which seismic activity may trigger these eruptions is still unclear. Could the self-accelerating exolution of gas observed in a beer bottle upon and impact be behind these geological phenomena as well?


\section{Adult bubbles: a life on the surface, to burst or not to burst}

Unless you are drinking bubbly on the space station or far away from any celestial object (see Box 3), bubbles will eventually move towards the free surface as a result of buoyancy. When the bubble reaches the surface, a liquid film will form on the exposed side of the bubble. This `surface' bubble will remain floating on the surface for some time, depending on the stability of the liquid film.
%
%

\newpage

\begin{center}
\colorbox{cyan}{\parbox{0.9\textwidth}{{\bf Box 3. Bubbles in microgravity}\\
It is interesting to think about how everyday things change in the absence of gravity. Bubbles provide an excellent example, in particular bubbles in drinks. Without gravity bubbles that grow in carbonated drinks are not pushed towards the surface; hence, they continue to grow to very large sizes \cite{Davison2007} resulting in a frothy liquid with a much larger gas volume fraction than its earth-equivalent. The image in figure \ref{fig:boiling} shows an example of this behaviour, for the case of boiling. The Movie 3 (in the supplemental material) shows another example of the differences in behaviour for the formation and growth of a bubbly cloud when gravity is absent. Interestingly enough, there have been some experiments aimed at producing carbonated soda for astronauts. There was a TV add in the 80s that advertised `Space Coke', which was an actual attempt from the famous cola drink to provide soda for the inhabitants of the space station. In addition to the large bubble sizes of space soda (that would also change the perception of flavor, as discussed above), the ingestion of such beverages would be problematic. The bubbles would not be able to escape the liquid within the digestive system, leading to painful bloating in the stomach and intestines. So, sorry, no bubbly drinks for space people!
}} 

\begin{figure}[ht!]
\centering
\includegraphics[width=0.9\textwidth]{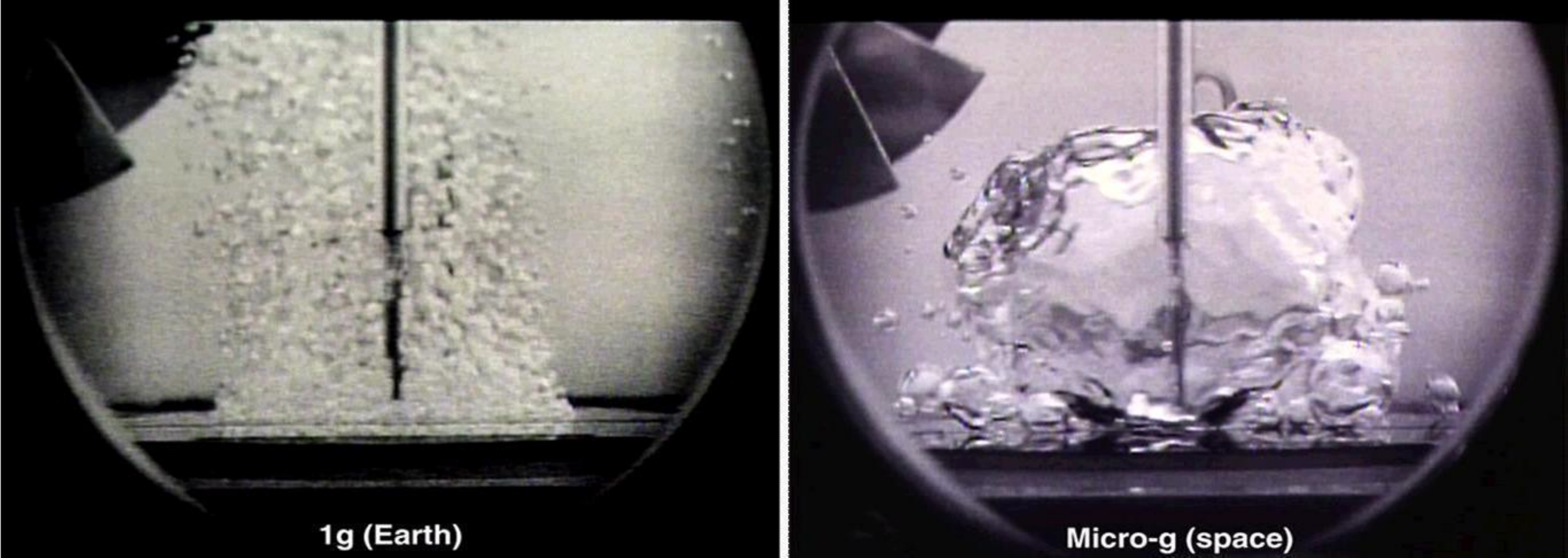}
\caption{Taken from report of the Nucleate Pool Boiling eXperiment (NPBX) in microgravity at International Space Station (for  more  details see Dhir et al. 
Microgravity Sci. Technol., 24:307, 2012) . An electric heater immersed in a liquid. When the liquid reaches its boiling temperature on earth, many small bubbles are formed and move upwards to the surface. Conversely, under the same conditions but in a micro-$g$ environment, a huge bubble form adjacent to the heater growing as large as the container. NOTE THAT THIS IMAGE GOES INSIDE BOX 3.}
\label{fig:boiling}
\end{figure}
\end{center}

\vspace{0.8cm}


How long would a bubble be able to remain on the surface is of great importance for drinks and, as it turns out, for many other interesting problems in nature and industry. The life time of a bubble is dictated by  the draining dynamics of the film, which progressively thins due to gravity until it ruptures. When the film is thin enough, about 10 \AA, van der Walls forces will become important and rupture will occur.

To understand bubble bursting, first we shall consider the shape and position of a bubble sitting on the surface. This state will be determined by the balance between buoyancy and surface tension forces\cite{Teixeira2015}, which for a bubble of size $R$, can be quantified by the Bond number, $Bo=\rho g R^2/\sigma=R^2/\lambda_c^2$, where $\sigma$ is the surface tension. This parameter compares the size of bubble with the so-called capillary length, $\lambda_c=\sqrt{\sigma/(\rho g)}$. For a small bubble, the value of the Bond number is small (considering the surface tension of clean water), which indicates that surface tension effects dominate over gravitational ones; hence, the bubble would remain mostly immersed in the liquid. On the other hand, larger bubbles (with correspondingly larger values of the Bond number) will protrude outside the surface level with a small meniscus. These differences can be observed in the image in figure \ref{fig:bubbly_drink}$e$. For the case of drinks, considering millimetric bubbles and aqueous properties,   the Bond number is of $O(0.1)$.

The process of film thinning is  dictated by a balance between gravity, viscosity and surface tension. A detailed modeling of the process is possible but rather elaborated since, in addition to solving the balances of mass and momentum of the liquid within the film coupled with the Young-Laplace equation (that relates the pressure across a curved interface with the surface tension), the thickness and position of the film evolve in time. However, by dimensional analysis we can deduce that:
\begin{equation}
    \frac{h}{h_o}=\Phi\left(\frac{h_o}{R}, \Pi_t, Bo\right)
\end{equation}
where $h$ is the thickness of the film ($h_o$ being the initial thickness) and $\Pi_t= t g R/\nu$ is a dimensionless time. Howell \cite{Howell1999} modelled this process analytically to obtain an expression for $h/h_o$ considering an asymptotic approximation for small $Bo$:
\begin{equation}
    \frac{h}{h_o}=\left(1+\frac{16}{\sqrt{3}}\frac{\Pi_t \sqrt{h_o/R}}{Bo^{3/2}}\right)^{-2}
\end{equation}
from which the rupture time, $t_{rup}$, can be calculated considering that at that time $h=h_{rup}$. For properties of water and a millimetric size bubble, $t_{rup}\sim 10^{-4}s$, considering $h_{rup}$ to be 10\AA. Clearly, this time is much shorter than what it is typically observed in any glass of a bubbly drink.

What is missing from the calculation above is the effect of impurities. Pure liquids, and specially water, are prone to accumulate molecules in their surfaces, the so-called surfactants. These molecules, which often have a head and a tail, one of which prefers to be in contact with air, accumulate at the interface causing several important effects. First, they can reduce the value of the surface tension. Soap is a classical example of this effect, being able to reduce the surface tension of water for up to three times. Increasing the value of $Bo$ by a factor of three does not significantly increase the rupture time. Most importantly, surfactant molecules tend to immobilize the surface. Therefore, the drainage would be subjected to a non-slip condition, significantly slowing down the process. For the specific case of alcoholic drinks, alcohol reduces the surface tension whereas the many other components in the beverages do contain surfactant molecules. For beer, it has been found that certain proteins, that are a by-product of the fermentation process, have a strong surfactant effect \cite{Blasco2011}. Hence, luckily, bubbles in beer last for long time periods. Furthermore, alcohol can also induce Marangoni flows; due to the different evaporation rate of alcohol and water, gradients of surface tension may appear in the surface of bubbles. Such gradients would, in turn, induce an upward-moving flow that would further delay the rupture process. This effect has been studied in detail for alcoholic drinks, known to cause the so-called ``tears of wine''.

When the bubble finally bursts, a complex motion of the liquid appears. As depicted in figure \ref{fig:bursting} (see also Movie 4, in the supplementary material), the tearing film retracts rapidly  leaving behind a curved cavity at the surface of the liquid. The surface tension tries to flatten the surface, moving quickly from a convex to a concave shape that leads to the formation of fast upwards jet. This effect is more notorious the smaller the bubble is, due to the larger capillary overpressure. Both the tearing film and the fragmentation of the jet lead to the formation of small aerosol droplets. This process has found to be of significant importance for the aerosol formation at the surface of the ocean\cite{Richter2016}; interestingly, it has also been found to significantly influence the emission of odors, which in turn affect the perception of flavor (see the study of Liger-Belair in champagne\cite{Liger2009}).
\begin{figure}[htp!]
\centering
\includegraphics[width=0.95\textwidth]{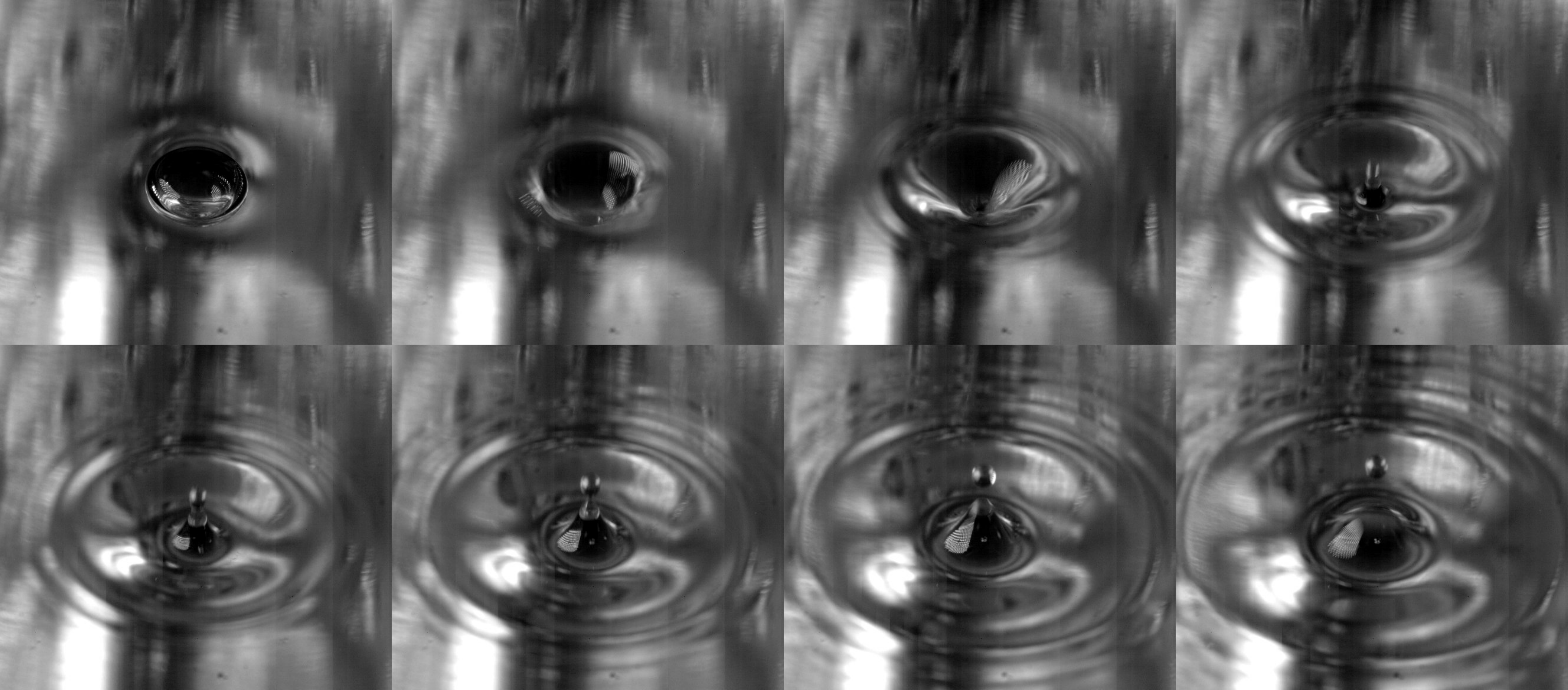}
\caption{Image sequence of a bubble bursting on the surface of water. A 2 mm, CO$_2$ bubble sitting on the surface of mineral water. When the film bursts, a strong upward jet is formed; the jet fragments to form small droplets. High speed camera photo sequence, taken at 1000 fps (taken by B. Palacios, UNAM).}
\label{fig:bursting}
\end{figure}

If bubbles do remain at the surface for some time, again as a result of physical properties of the interface, it is possible for them to accumulate. If the production rate of bubbles in the bulk of the liquid is faster than the life time on the surface, then a foam head forms. This region of densely packed bubbles that sits on the surface is a very important characteristic for many drinks and it serves a particular purpose: it contributes to release odors which affect the perception of flavor and to yield some beers a creamy mouthfeel. The thickness of the foam layer depends on the liquid properties (due to the drainage time of the film) and, as we all know, on  the skill of the person who pours the bubbly liquid from the bottle or tap. If the stream of liquid is too fast,  air is entrained into the liquid. Such air cavities would serve as nucleation sites (see section \ref{sec:birth_rise}) that, in turn, would increase the production rate of bubbles. If the expected life time of bubbles is not affected, then the thickness of the foam layer would increase, sometimes, sadly, overcoming the rim of the glass and spilling the bubbly treasure. How much foam is good? Depends on who you ask and where do they come from! A recent study suggests that having a layer of foam reduces the sloshing in a vibrated container\cite{Viola2016}; so, if you prefer not to spill your beer while walking, it is better to have it with foam. It has also been shown that foam slows down degassing of beer and also serves as a thermal insulator. Some beers served in the LA Dodger stadium has an additional `artificial' foam; it helps to keep your drink cool for longer times!

\begin{figure}[h!]
\centering
\centering
    \begin{tabular}{cc}
    \includegraphics[height=0.3\textwidth]{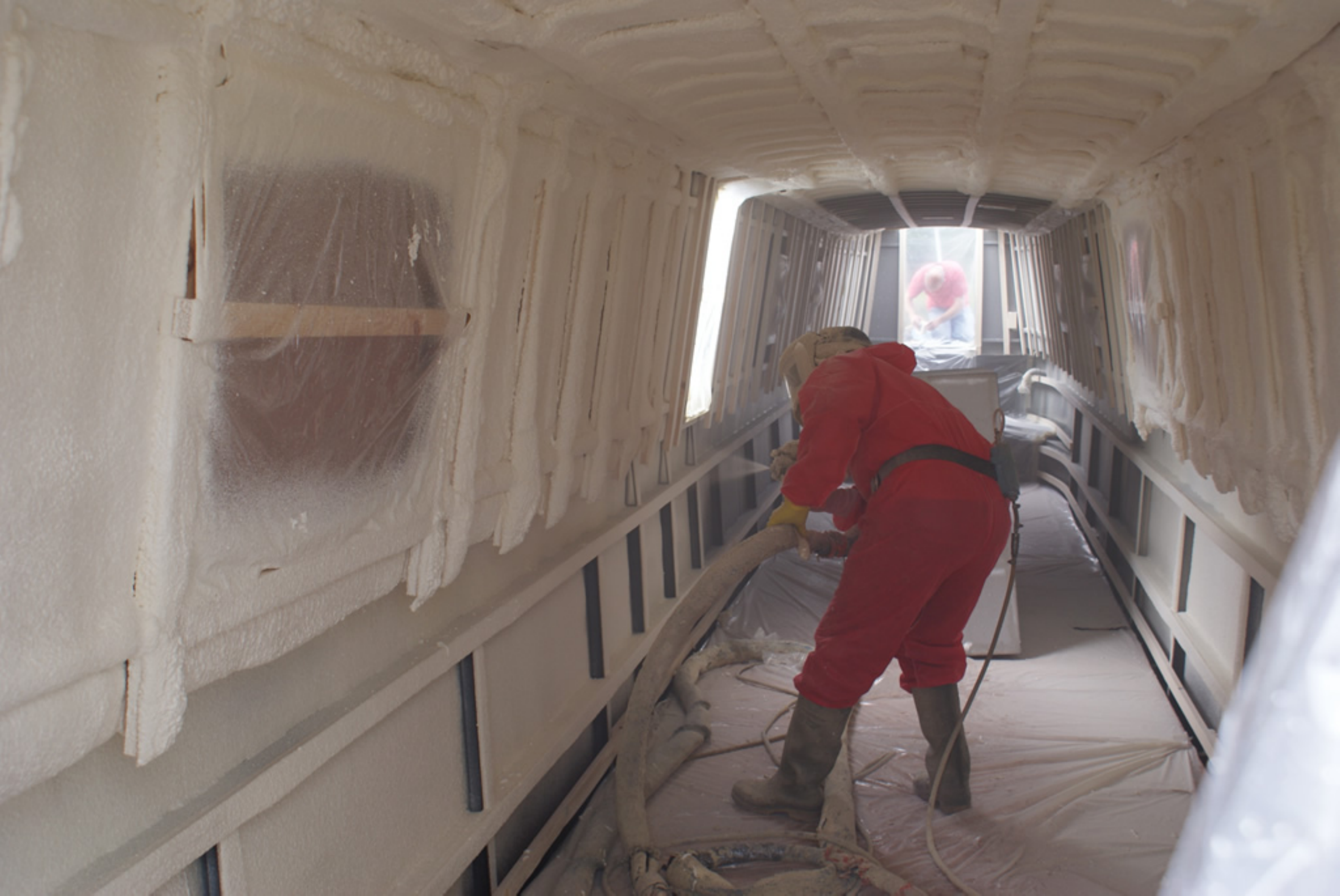} &
    \includegraphics[height=0.3\textwidth]{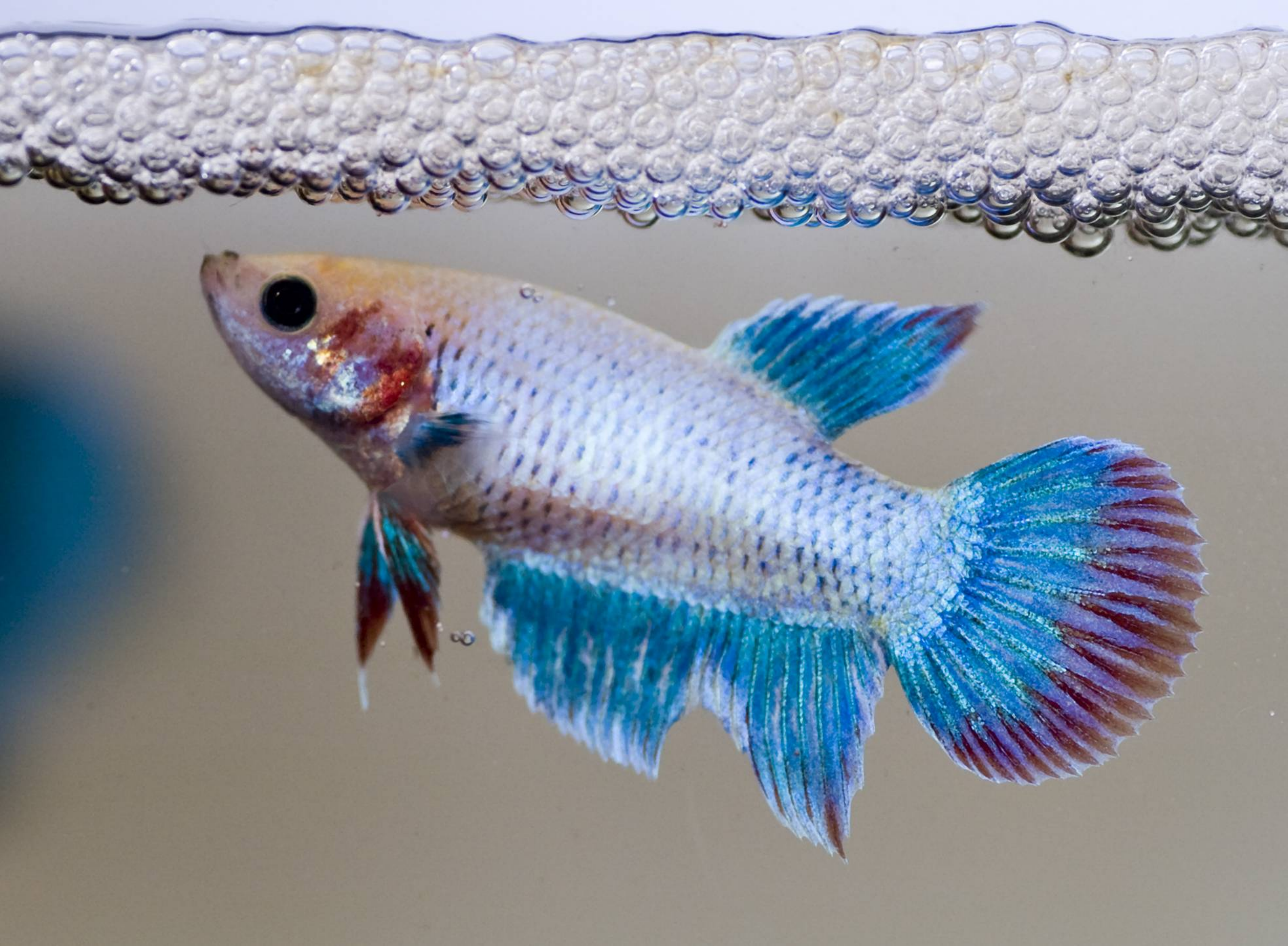}\\
    ($a$) & ($b$) \\
    \includegraphics[height=0.3\textwidth]{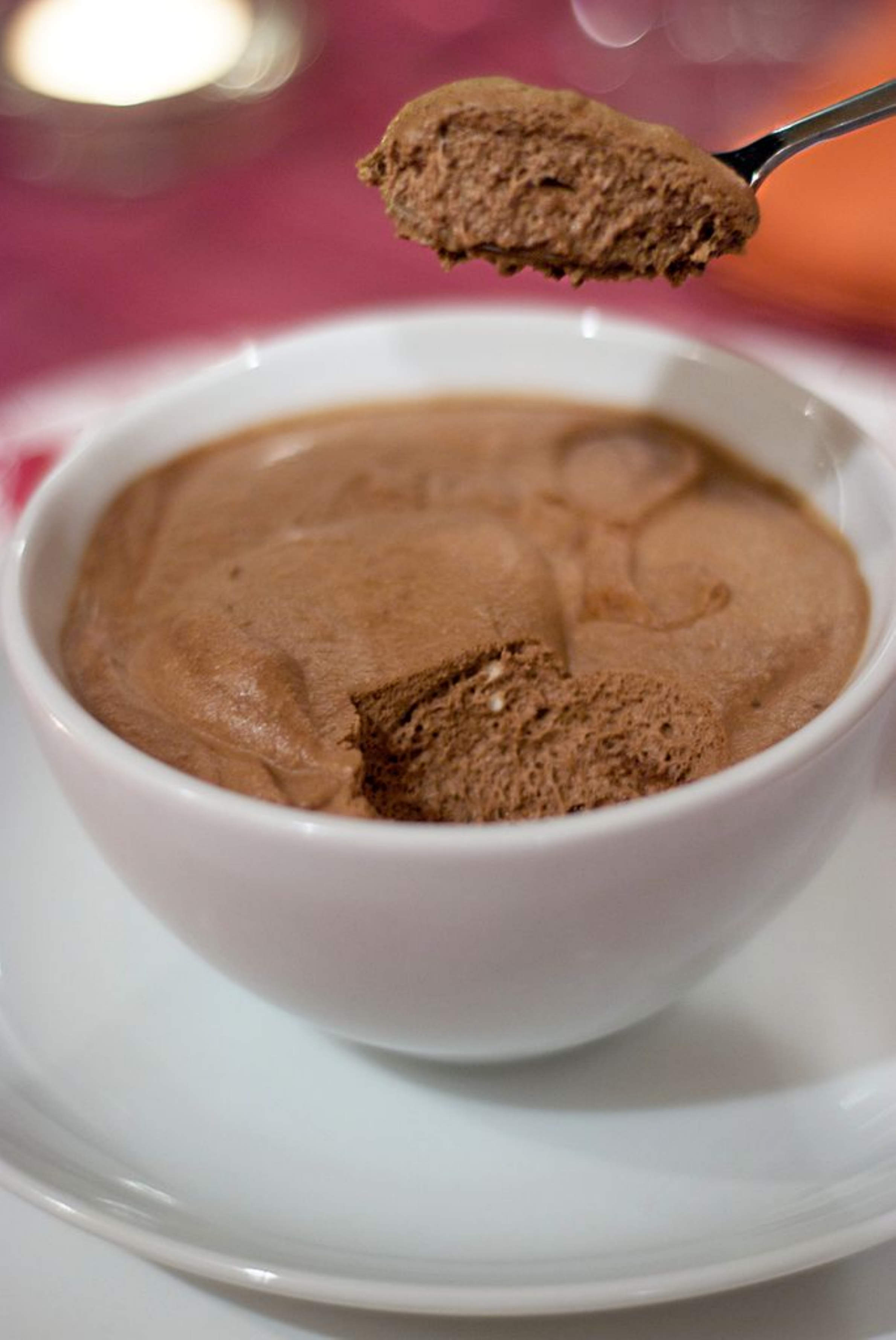} &
    \includegraphics[height=0.3\textwidth]{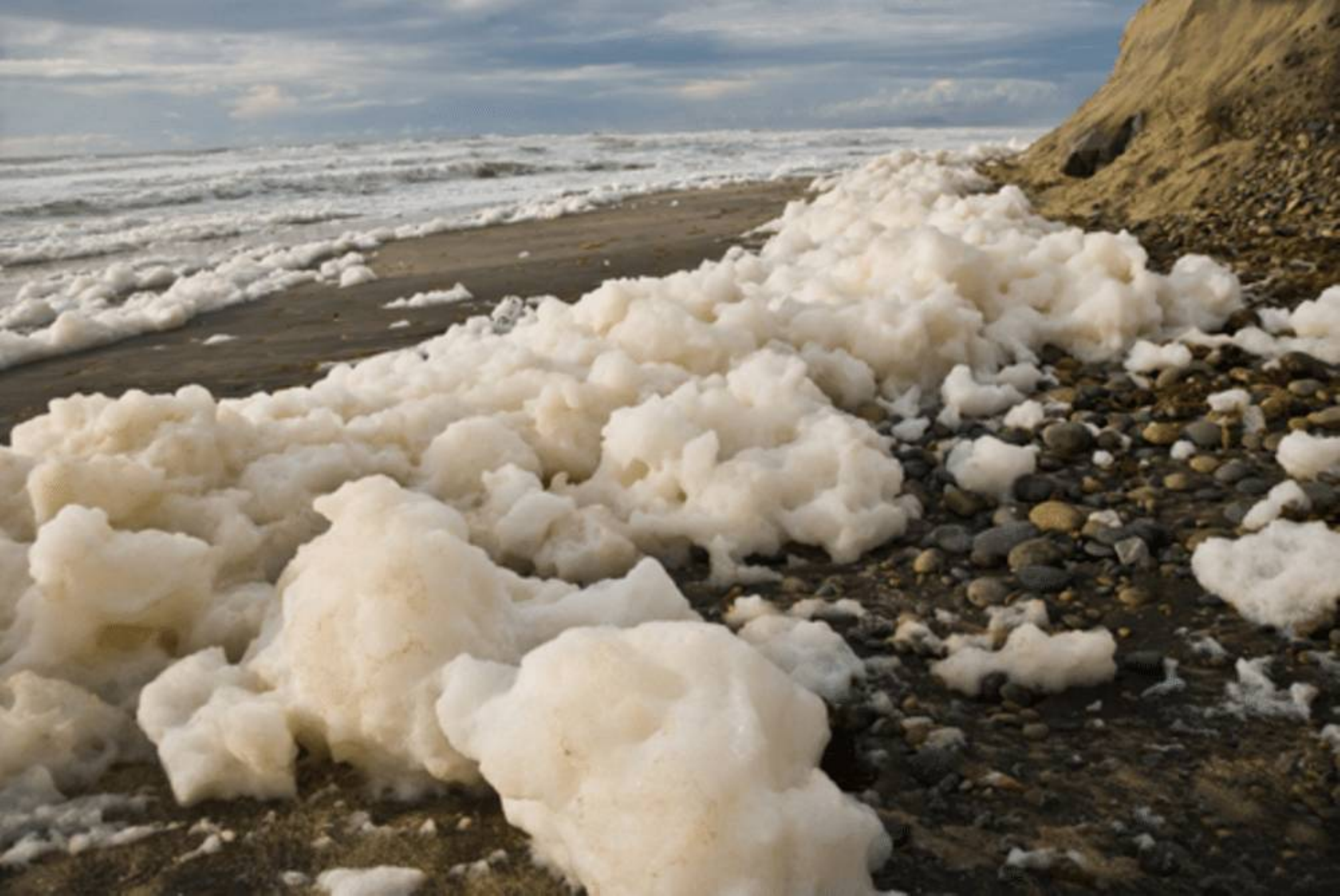}\\
    ($c$) & ($d$)
    \end{tabular}
\caption{Foams everywhere: (a) foam is applied inside walls improve their thermal insulation properties (Photo courtesy of Isothane industries); (b) the nest of Siamese Beta fish is a foam. It keeps the eggs in an oxygen rich environment and protected from thermal fluctuations (photo from Pintrest, CC license); (c) chocolate mousse would not be the same if it did not have bubbles (photo from wikimedia commons, CC license); (d) Sea foam in the coast of California after a storm (photo by G. Jorda, CC license 2010).}
\label{fig:foams}
\end{figure}

The properties of foam have been studied widely\cite{Cohen2013} since, in addition of adding appeal to a drink, it has important uses in numerous fields as depicted in figure \ref{fig:foams}. Here we list just a few. Some of the important properties of foam are its stability (basically its life time), size distribution and mechanical properties. Foam can be viscoelastic and have a bulk shear-dependent viscosity. The thermal conductivity of foam is very small, a property that makes this material and excellent insulator. In fact, most modern insulating materials are foam-based products (see the image in figure \ref{fig:foams}$a$) Another important industrial, currently relevant, use of this material is in oil recovery. Foam is injected into fractured wells to  plug small pores and force  the oil to flow in larger recovery channels.  For biological systems water-soluble proteins can significantly prolong the life time of foam. Some fishes and amphibians create foam nests to deposit and hatch their eggs\cite{Jaro2001}, as shown in figure \ref{fig:foams}$b$. These nests secure the eggs from temperature fluctuations and provide a oxygen-rich environment that favors hatching. For the food industry, foam is very important as it adds texture to other-wise boring food\cite{Dickinson2015}. The best example of edible foam is \emph{mousse au chocolat}, deliciously shown in \ref{fig:foams}$c$. The proteins in egg white and sugar can make foam to last for long time periods. The foamy texture is what endows this dessert its appeal. Cappuccino would be rather plain without milk foam, which is stable also because of proteins and fat. Lastly, the foam on the surface of the sea water is usually an indication of the presence of dissolved organic matter\cite{Schilling2011}. It is sometimes an indication of pollution or contamination. The image in figure \ref{fig:foams}$d$ shows a typical example. The tracking of white-cap waves is a widely used tool to measure wind speed over the ocean via satellite imaging. Such measurement could not be conducted in the sea-water foam were not sufficiently stable.

\section{Conclusions}

We hope that, after reading this article, the next time that you will sit before a glass of a bubbly drink your will look at it in a different way: as a laboratory where a myriad of fascinating phenomena take place, with length scales ranging from tens of microns (where bubbles are born at nucleation sites) to tens of centimeters (the container).

This range of scales is swept thanks to the strong mass transfer occurring between the bubbles and the drink, which makes them grow substantially. This is due to two reasons: on the one hand, the high solubility of CO$_2$ in water and on the other hand to the two-way coupling between bubble growth and rise and the large-scale circulation that their upwards motion induces in the glass. This two-way coupling lies at the heart of the fast degassing of carbonated drinks. We argue that similar phenomena are observed in several geophysical flows and the question remains open of whether other phenomena, such as the triggering of volcanic eruptions by remote earthquakes, are also connected to the self-accelerating nature of bubbly plumes.

Once bubbles reach the free surface two scenarios can occur: if the arrival rate is faster than the bubble bursting time, a foam head forms, beer being an excellent example of this. Contrarily, if bubbles burst faster than they arrive at the surface, then a clean surface forms, with bubble bursting serving as a source of a fine spray that enhances the drink taste and odor, as happens in champagne. Either way, surfactants play a central role either in stabilizing the foam, or in delaying its burst thus making it more intense.

In summary, carbonated beverages are portable laboratories that can be used to demonstrate in an amusing way the working of many flows also found in nature and industry. So much for a glass of a refreshing bubbly drink.\\

\newpage
{\it We are indebted to Almudena Casado for providing us with the experimental data used in figure \ref{fig:bubble_rise}.
J.R.-R. acknowledges the support of the Spanish Ministry of Economy and Competitiveness through grants DPI2015-71901-REDT and DPI2017-88201-C3-3-R, partly funded through European Funds. R.Z. acknowledges the continued support of UNAM, CONACyT-Mexico and PEMEX over the years to explore different aspects of bubbly flows.}

%
%

\bibliographystyle{unsrt}

\bibliography{references}

\newpage

\appendix

\section*{Supplementary material}

\begin{itemize}
\item Cover candidate photo. Bubbles, foam and aerosol droplets in a freshly poured glass of apple soda. File: \href{https://drive.google.com/open?id=1aVUzne4VbYCNfjsrhh-CfsIBjGIIBspn}{\texttt{Bubbles\_drinks\_cover.jpg}}. See full resolution in tiff format here: \href{https://drive.google.com/open?id=1DK89-y5lJLNz92ASX25e_wNTZzON1Yo_}{file}. 
Photo credit: R. Vi\~nas, TresArt Collective, 2018.

\item Movie 1. Apple soda being served into a glass. This movie illustrates some of the phenomena depicted in figure \ref{fig:bubbly_drink}, in particular the rise of bubbles to the surface and the processes that they undergo there: bursting and, when the bubble arrival rate is large, agglomeration. Taken at 240 fps, played at 60 fps, Iphone 6, by R. Zenit. File: \href{https://drive.google.com/open?id=1o8oHs3jyWZC0TcEEl0X98tkCTSEp7uDy}{\texttt{Bubbles\_drinks\_1.mov}}.

\item Movie 2. Bubbles forming at a cellulose fiber immersed in lager beer, with a bottling pressure around 5 bar. This movie is a close-up to the origin of a bubble trail like those illustrated in figure \ref{fig:bubbly_drink}$b$. A gas cavity trapped inside the fiber (see panel $c$) serves as a source of bubbles that, when reaching a size large enough, rise to the surface close to the glass wall. The length of the fiber is around 1 mm. File: \href{https://drive.google.com/open?id=17Uc1N2aamg0JX6hsyhWrwHHbHB4ck_Xt}{\texttt{Bubble\_drinks\_2.mov}} .

\item Movie 3. Bubbly cloud in microgravity. This movie compares the evolution of bubble clouds generated by a spark inside carbonated water. The left movie shows a bubble cloud growing in normal gravity conditions whereas the right one corresponds to a bubble cloud growing under microgravity conditions. These experiments were performed in the drop tower at ZARM, Bremen, Germany (For more information, see \href{https://arxiv.org/abs/1703.08875}{Patricia Vega-Mart\'{\i}nez {\it et al.} {\it Microgravity Science and Technology}, {\bf 29}, pp 297-304, 2017}). File: \href{https://drive.google.com/open?id=1tVTPdGa45YxRd4Sj7bxSFfLcYsjhvlXq}{\texttt{Bubbles\_drinks\_3.mov}}.

\item Movie 4. The bursting of a superficial bubble. Carbon dioxide bubble on top of mineral water. Taken at 1000 fps, played at 5 fps. Taken by B. Palacios, UNAM. File: \href{https://drive.google.com/open?id=10z9jzcKi_nRWS86gkDpzlIs2q-IvUP89}{ \texttt{Bubbles\_drinks\_4.avi}}.

\end{itemize}

\newpage

\section*{Source images for Figure 1}
The images are not cropped.
\begin{itemize}

\item Background image. A glass full of beer. \href{https://drive.google.com/open?id=1FvfkhfZ9csnVkq0Gp_jrvgDDoXAWPKKH}{file}.\\
Photo by: R. Vi\~nas, TresArt Collective. 

\item Panel a. Foam interface. 
\href{https://drive.google.com/open?id=1MomcO13nb1ugJ5zRb_1cbgbtSVHmtsxM}{file}.\\
Photo by: R. Vi\~nas, TresArt Collective. 

\item Panel b. Bubble chain. 
\href{https://drive.google.com/open?id=1MQQRPQhmG9P0jGdSBc_WP21UyIr8FEdH}{file}.\\
Photo by: R. Vi\~nas, TresArt Collective. 

\item Panel c. Bubble birth at a cellulose fiber. 
\href{https://drive.google.com/open?id=1tOlBRfYcXbTzTbKzxCeTi4C7vtZDTPUV}{file}.\\
Photo by: J. Rodriguez, UC3-M. 

\item Panel d. Bubble cluster after tapping. 
\href{https://drive.google.com/open?id=15QgxOIGDgPnIbD-uk74xmuVqazHY3lfs}{file}.\\
Photo by: J. Rodriguez, UC3-M. 

\item Panel d. Surface bubbles. 
\href{https://drive.google.com/open?id=1LxfBSKttrrywBAhTyjSNFoXseN3NqpvN}{file}.\\
Photo by: R. Zenit, UNAM.

\end{itemize}

\end{document}